\documentclass[aps,pra]{revtex4}
\usepackage{graphicx}

\begin{document}

\title{Improving the security of quantum direct communication with authentication
\thanks{Email: zjzhang@ahu.edu.cn}}
\author{Zhan-jun Zhang \\
{\normalsize School of Physics \& Material Science, Anhui
University, Hefei 230039, Anhui, China}\\
{\normalsize Email address: zjzhang@ahu.edu.cn }}

\date{\today}
\maketitle

\begin{minipage}{430pt}
Two protocols of quantum direct communication with authentication
[Phys. Rev. A {\bf 73}, 042305 (2006)] are recently proposed by
Lee, Lim and Yang. In this paper we will show that in the two
protocols the authenticator Trent should be prevented from knowing
the secret message of communication. The first protocol can be
eavesdropped by Trent using the the intercept-measure-resend
attack, while the second protocol can be eavesdropped by Trent
using single-qubit measurement. To fix these leaks, I revise the
original versions of the protocols by using the Pauli-Z operation
$\sigma_z$ instead of the original bit-flip operation $X$. As a
consequence, the protocol securities are improved.  \\

\end{minipage}\\

Quantum key distribution (QKD) is one of the most interesting
topics in quantum information processing, which provides a novel
way for two legitimate parties to share a common secret key over a
long distance with negligible leakage of information to an
eavesdropper Eve. Its ultimate advantage is the unconditional
security. Hence, after Bennett and Brassard's pioneering work
published in 1984[1], much attentions have been focused on this
topic and a variety of quantum communication protocols[1-13,17-18]
have been proposed. In these works, various properties of quantum
mechanics, such as no-cloning theorem, uncertainty principle,
entanglement, indistinguishability of nonorthogonal states,
non-locality, and so on, are used to accomplish QKD tasks.
Different from QKD, the deterministic quantum secure direct
communication (QSDC) protocol is to transit directly the secret
messages without first generating QKD to encrypt them. Hence it is
very useful and usually desired, especially in some urgent time.
However, a deterministic secure direct communication protocol is
more demanding on the security. Therefore, only recently a few of
deterministic secure direct communication protocols have been
proposed[3-12] and some of them are essentially insecure[13-15].
Two of the QSDC protocls are the Lee-Lim-Yang protocols of quantum
direct communication with authentication[12] proposed very
recently by Lee, Lim and Yang using the Greenberg-Horne-Zeilinger
(GHZ) states[16]. Based on some security analysis Lee, Lim and
Yang claimed that their two protocols are secure. However, in this
paper we will show that in the two protocols the authenticator
Trent should be prevented from knowing the secret message of
communication. The first protocol can be eavesdropped by Trent
using the the intercept-measure-resend attack, while the second
protocol can be eavesdropped by Trent using single-qubit
measurement. I will fix these leaks by modifying the original
versions of the protocols using that Pauli-Z operation $\sigma_z$
instead of the original bit-flip operation $X$ so that the
protocol securities are improved.

There are three parties in each of the Lee-Lim-Yang protocols.
Alice and Bob are the legitimate users of the communication. Trent
is the third party who is introduced to authenticate the two users
participating in the communication. Trend is assumed to be more
powerful than the other two parties and he supplies the GHZ states
each in the form of
$|\Psi\rangle=(|000\rangle+|111\rangle)/\sqrt{2}$. The protocols
are composed of two parts: one is for an authentication process
and the other is for a direct communication. The authentication
process is same for both  Lee-Lim-Yang protocols. After the
authentication, there are two possibilities for Alice to send
qubits: one is to Bob and the other is to Trend. The former case
corresponds to the Lee-Lim-Yang protocol 1 and the latter case to
the Lee-Lim-Yang protocol 2. This is a difference between the two
protocols. Incidentally, there is an unphysical mistake about the
respondences in the text of Ref.[12].

The purpose of the authentication process in the Lee-Lim-Yang
protocols is to let the three participants safely share GHZ
states. To achieve this goal, it is assumed that Trend should
share in priori secret authentication keys $K_{ta}$ and $K_{tb}$
with Alice and Bob, respectively. The lengths of the
authentication keys $K_{ta}$ and $K_{tb}$ are larger than the
length of the bit string of the secret message which will be
communicated from Alice to Bob. According to the one-time pad
cryptography, when the private key length is equal to the secret
message length, the secret message can be securely communicated to
remote places after encryption. If Trend is not assumed to prevent
from knowing the secret message, then in this case the secret
message can be transferred in the following very simple classical
way instead of using the Lee-Lim-Yang QSDC protocols, i.e, Alice
can securely send the secret message to Trend by using their
secret authentication key $K_{ta}$ and then Trend can securely
send the secret message to Bob by using their secret
authentication key $K_{tb}$. Hence, in the Lee-Lim-Yang protocols
the third party Trend should be prevented from knowing the secret
message though he is introduced to authenticate the communication.

Assume that the GHZ states are safely shared among the three
parties after the authentication process. Let us now briefly
review the second part of the Lee-Lim-Yang protocol 1.

(a) Alice selects a subset of GHZ states of her remaining set
after authentication and keeps it secret.

(b) Alice chooses a random bit string which has no correlation to
the secret message to transmit to Bob. This bit string will be
used to check the security of the channel.

(c) Following the random bit string, Alice performs unitary
operations on the the qubits selected for check process. The
unitary operations are defined as follows. The bit '0' correspond
to $H$ and the bit '1' to $HX$, where $H$ is the Hadamard
operation and $X$ is the bit-flip operation. The GHZ states after
Alice's operations are transformed into:
\begin{eqnarray}
H_A|\Psi\rangle &=& H_A(|000\rangle_{ATB}+|111\rangle_{ATB})/\sqrt{2} \nonumber \\
&=&\frac{1}{2}\{|000\rangle_{ATB}+|100\rangle_{ATB}+|011\rangle_{ATB}-|111\rangle_{ATB}\}
\nonumber \\
&=&\frac{1}{2}\{(|\phi^+\rangle_{AB}-|\psi^-\rangle_{AB})|-\rangle_T
+(|\phi^-\rangle_{AB}+|\psi^+\rangle_{AB})|+\rangle_T\} , \\
H_AX_A|\Psi\rangle &=& H_A(|100\rangle_{ATB}+|011\rangle_{ATB})/\sqrt{2} \nonumber \\
&=&\frac{1}{2}\{|000\rangle_{ATB}-|100\rangle_{ATB}+|011\rangle_{ATB}+|111\rangle_{ATB}\}
\nonumber \\
&=&\frac{1}{2}\{(|\phi^-\rangle_{AB}-|\psi^+\rangle_{AB})|-\rangle_T
+(|\phi^+\rangle_{AB}+|\psi^-\rangle_{AB})|+\rangle_T\},
\end{eqnarray}
where $|\phi^\pm\rangle=(|00\rangle\pm|11\rangle)/\sqrt{2}$,
$|\psi^\pm\rangle=(|01\rangle\pm|10\rangle)/\sqrt{2}$ and
$|\pm\rangle=(|0\rangle\pm|1\rangle)/\sqrt{2}$.

(d) Alice encodes the secret message with a classical error
correction code on the remaining GHZ states in terms of the
unitary operation definition in (c).

(e) After making all unitary operations, Alice sends the encoded
qubits to Bob.

(f) Bob makes Bell measurements on the pairs of particles
consisting of his qubits and Alice's qubits.

(g) Trent measures his third qubit in the $x$ basis
$\{|-\rangle,|+\rangle\}$ and publishes the measurement outcomes.

(h) Using Trend's measurement outcomes and his Bell-state
measurement outcomes, Bob infers Alice's secret bits consisting of
both the random bits and the secret message.

(i) Bob lets Alice reveal the check bits' positions and values.

(j) Bob can know whether the channel is disturbed according to the
error rate. If the error rate is higher than expected, an
eavesdropper is concluded in the communication but fortunately the
secret message is not leaked out. If the error rate is lower, Bob
can extract the secret message (see Table I in Ref.[12]).

As I show before, the authenticator Trent should be prevented from
the secret message. Otherwise, the communication can be realized
in a very simple classical way. Although the Lee-Lim-Yang protocol
1 is claimed to be secure (that is, the secret message can not be
leaked out), in the protocol 1 the insider Trent can eavesdrop the
secret message by using the intercept-measure-resend attack. This
can be seen as follows. When Alice sends her encoded qubits to
Bob, Trent intercepts the qubits and performs $H$ operation on
each qubit. In this case, the whole states are transformed into
\begin{eqnarray}
H_AH_A|\Psi\rangle &=& (|000\rangle_{ATB}+|111\rangle_{ATB})/\sqrt{2},\\
H_AH_AX_A|\Psi\rangle &=&
(|100\rangle_{ATB}+|011\rangle_{ATB})/\sqrt{2}.
\end{eqnarray}
After the unitary operations, Trent measures Alice's qubit and his
qubit in the $z$ basis $\{|0\rangle,|1\rangle\}$, respectively. If
the two outcomes are same, then Trent can conclude that Alice has
performed a $H$ operation corresponding to the bit '0'. Otherwise,
Alice has performed a $HX$ operation corresponding to the bit '1'.
In this case, Trent has already known for each Alice's qubit which
unitary operation Alice has performed on. Alternatively, he has
already got Alice's whole bit string including both the random bit
string for check and the secret message. In the following what he
needs to do is to remove the random bits. Fortunately, in the step
(i), Alice will publish which qubits are used as check qubits.
This means that the authenticator Trent can completely know the
secret message using this intercept-measure-resend attack.

After Trnent's attack, he sends Alice's qubits to Bob. One can
easily find that for Alice an Bob the error rate will obviously be
higher than expected. Unfortunately, Alice and Bob only know the
channel is disturbed and still believe that the secret message is
not leaked out.

To fix this leak, we think the original version of Lee-Lim-Yang
protocol 1 can be modified by using the Pauli-Z operation
$\sigma_z$ instead of the original bit-flip operation $X$. In this
case, the total states after Alice's operations are represented as
follows:
\begin{eqnarray}
H_A|\Psi\rangle &=& H_A(|000\rangle_{ATB}+|111\rangle_{ATB})/\sqrt{2} \nonumber \\
&=&\frac{1}{2}\{|000\rangle_{ATB}+|100\rangle_{ATB}+|011\rangle_{ATB}-|111\rangle_{ATB}\}
\nonumber \\
&=&\frac{1}{2}\{(|\phi^+\rangle_{AB}-|\psi^-\rangle_{AB})|-\rangle_T
+(|\phi^-\rangle_{AB}+|\psi^+\rangle_{AB})|+\rangle_T\}, \\
H_A\sigma_{zA}|\Psi\rangle &=& H_A(|000\rangle_{ATB}-|111\rangle_{ATB})/\sqrt{2} \nonumber \\
&=&\frac{1}{2}\{|000\rangle_{ATB}+|100\rangle_{ATB}-|011\rangle_{ATB}+|111\rangle_{ATB}\}
\nonumber \\
&=&\frac{1}{2}\{(|\phi^-\rangle_{AB}+|\psi^+\rangle_{AB})|-\rangle_T
+(|\phi^+\rangle_{AB}-|\psi^-\rangle_{AB})|+\rangle_T\}.
\end{eqnarray}
After this modification, if Trend intercepts Alice's encoded
qubits and performs $H$ operations, then the total states are
transformed into
\begin{eqnarray}
H_AH_A|\Psi\rangle &=& (|000\rangle_{ATB}+|111\rangle_{ATB})/\sqrt{2},\\
H_AH_A\sigma_{zA}|\Psi\rangle &=&
(|000\rangle_{ATB}-|111\rangle_{ATB})/\sqrt{2}.
\end{eqnarray}
If he measures respectively his qubit and Alice's qubit in the $z$
basis, the outcomes will always be same. In this case, he can not
know for each of Alice's qubits which unitary operation Alice has
performed on. This means Trent can not know Alice's secret
message. However, one can easily find that revised protocol works
successfully as for as Alice and Bob's communication is concerned.
See Table 1 for a brief summary.

\begin{center}
Table 1 The relation of Alice's operation, Bob's measureemnt, and
Trent's announcement in the revised Lee-Lim-Yang protocol 1 can be
summarized as follows.
\begin{tabular}{ccccccccc} \hline
Trent's publication &&& Bob's Measurement &&& Alice's operation \\
\hline $|+\rangle_T$ &&& $|\phi^+\rangle_{AB}$ or
$|\psi^-\rangle_{AB}$&&&  $H\sigma_z \ \ (1)$\\  $|+\rangle_T$
&&&
$|\phi^-\rangle_{AB}$ or $|\psi^+\rangle_{AB}$&&&  $H \ \ \ \ \ (0)$\\
$|-\rangle_T$ &&& $|\phi^+\rangle_{AB}$ or
$|\psi^-\rangle_{AB}$&&& $H \ \ \ \ \ (0)$\\  $|-\rangle_T$  &&&
$|\phi^-\rangle_{AB}$ or $|\psi^+\rangle_{AB}$&&&  $H\sigma_z \ \ (1)$\\
\hline
\end{tabular}\\
\end{center}

\vskip 0.5cm

Let us now briefly review the second part of the Lee-Lim-Yang
protocol 2. Some steps are same for both protocols. Nevertheless,
for completeness, I depict all the steps as follows.

($a'$) Alice selects a subset of GHZ states of her remaining set
after authentication and keeps it secret.

($b'$) Alice chooses a random bit string which has no correlation
to the secret message to transmit to Bob. This bit string will be
used to check the security of the channel.

($c'$) Following the random bit string, Alice performs unitary
operations on the the qubits selected for check process.  The GHZ
states after Alice's operations are transformed into:
\begin{eqnarray}
H_A|\Psi\rangle &=& H_A(|000\rangle_{ATB}+|111\rangle_{ATB})/\sqrt{2} \nonumber \\
&=&\frac{1}{2}\{(|\phi^+\rangle_{AT}-|\psi^-\rangle_{AT})|-\rangle_B
+(|\phi^-\rangle_{AT}+|\psi^+\rangle_{AT})|+\rangle_B\}, \\
H_AX_A|\Psi\rangle &=& H_A(|100\rangle_{ATB}+|011\rangle_{ATB})/\sqrt{2} \nonumber \\
&=&\frac{1}{2}\{(|\phi^-\rangle_{AT}-|\psi^+\rangle_{AT})|-\rangle_B
+(|\phi^+\rangle_{AT}+|\psi^-\rangle_{AT})|+\rangle_B\}.
\end{eqnarray}

($d'$) Alice encodes the secret message with a classical error
correction code on the remaining GHZ states in terms of the
unitary operation definition in ($c'$).

($e'$) After making all unitary operations, Alice sends the
encoded qubits to Trent.

($f'$) Trent makes Bell measurements on the pairs of particles
consisting of his qubits and Alice's qubits and publishes the
measurement outcomes.

($g'$) Bob measures his qubits in the $x$ basis
$\{|-\rangle,|+\rangle\}$ .

($h'$) Using Trend's Bell-state measurement outcomes and his
measurement outcomes, Bob infers Alice's secret bits consisting of
both the random bits and the secret message.

($i'$) Bob lets Alice reveal the check bits' positions and values.

($j'$) Bob can know whether the channel is disturbed according to
the error rate. If the error rate is higher than expected, an
eavesdropper is concluded in the communication but fortunately the
secret message is not leaked out. If the error rate is lower, Bob
can extract the secret message (see Table II in Ref.[12]).

In the protocol 2 the insider Trent can eavesdrop the secret
message by using the measurement attack as follows. Trent performs
$H$ operation on each qubit he received from Alice. In this case,
the whole states are transformed into
\begin{eqnarray}
H_AH_A|\Psi\rangle &=& (|000\rangle_{ATB}+|111\rangle_{ATB})/\sqrt{2},\\
H_AH_AX_A|\Psi\rangle &=&
(|100\rangle_{ATB}+|011\rangle_{ATB})/\sqrt{2}.
\end{eqnarray}
After the unitary operations, Trent measures Alice's qubit and his
qubit in the $z$ basis $\{|0\rangle,|1\rangle\}$, respectively. If
the two outcomes are same, then Trent can conclude that Alice has
performed a $H$ operation corresponding to the bit '0'. Otherwise,
Alice has performed a $HX$ operation corresponding to the bit '1'.
In this case, Trent has already known for each Alice's qubit which
unitary operation Alice has performed on. Alternatively, he has
already got Alice's whole bit string including both the random bit
string for check and the secret message. In the following what he
needs to do is to remove the random bits. Fortunately, in the step
(i'), Alice will publish which qubits are used as check qubits.
This means that the authenticator Trent can completely know the
secret message using this measurement attack.

After Trnent's attack, he randomly publishes his measurement
outcomes. One can easily find that for Alice an Bob the error rate
will obviously be higher than expected. Unfortunately, Alice and
Bob only know the channel is disturbed and still believe that the
secret message is not leaked out.

To fix this leak, the original version of Lee-Lim-Yang protocol 2
can also be modified by using the Pauli-Z operation $\sigma_z$
instead of the original bit-flip operation $X$. In this case, the
total states after Alice's operations are represented as follows:
\begin{eqnarray}
H_A|\Psi\rangle &=& H_A(|000\rangle_{ATB}+|111\rangle_{ATB})/\sqrt{2} \nonumber \\
&=&\frac{1}{2}\{(|\phi^+\rangle_{AB}-|\psi^-\rangle_{AB})|-\rangle_T
+(|\phi^-\rangle_{AB}+|\psi^+\rangle_{AB})|+\rangle_T\}, \\
H_A\sigma_{zA}|\Psi\rangle &=&
H_A(|000\rangle_{ATB}-|111\rangle_{ATB})/\sqrt{2}
\nonumber \\
&=&\frac{1}{2}\{(|\phi^-\rangle_{AT}+|\psi^+\rangle_{AT})|-\rangle_B
+(|\phi^+\rangle_{AT}-|\psi^-\rangle_{AT})|+\rangle_B\}.
\end{eqnarray}
After this modification, if Trend intercepts Alice's encoded
qubits and performs $H$ operations, then the total states are
transformed into
\begin{eqnarray}
H_AH_A|\Psi\rangle &=& (|000\rangle_{ATB}+|111\rangle_{ATB})/\sqrt{2},\\
H_AH_A\sigma_{zA}|\Psi\rangle &=&
(|000\rangle_{ATB}-|111\rangle_{ATB})/\sqrt{2}.
\end{eqnarray}
If he measures respectively his qubit and Alice's qubit in the $z$
basis, the outcomes will always be same. In this case, he can not
know for each of Alice's qubits which unitary operation Alice has
performed on. This means Trent can not know Alice's secret
message. However, one can easily find that revised protocol works
successfully as for as Alice and Bob's communication is concerned.
See Table 2 for a brief summary.

\begin{center}
Table 2 The relation of Alice's operation, Bob's measureemnt, and
Trent's announcement in the revised Lee-Lim-Yang protocol 2 can be
summarized as follows.
\begin{tabular}{ccccccccc} \hline
Trent's announcement &&& Bob's Measurement &&& Alice's operation \\
\hline 0 \ ($|\phi^+\rangle_{AB}$ or $|\psi^-\rangle_{AB}$) &&&
$|+\rangle_T$ &&&  $H\sigma_z \ \ (1)$\\   0 \
($|\phi^+\rangle_{AB}$ or $|\psi^-\rangle_{AB}$) &&&
$|-\rangle_T$ &&&  $H \ \ \ \ \ (0)$\\
1 \ ($|\phi^-\rangle_{AB}$ or $|\psi^+\rangle_{AB}$) &&&
$|+\rangle_T$ &&& $H \ \ \ \ \ (0)$\\  1 \ ($|\phi^-\rangle_{AB}$
or $|\psi^+\rangle_{AB}$) &&& $|-\rangle_T$
&&&  $H\sigma_z \ \ (1)$\\
\hline
\end{tabular} \\
\end{center}

\vskip 0.5cm

To summarize, in this paper we have shown that the Lee-Lim-Yang
protocols can be eavesdropped by the authenticator Trent using
some specific attacks and we have revised the original versions of
the protocols by using the Pauli-Z operation $\sigma_z$ instead of
the original bit-flip operation $X$ so that
the protocol securities are improved.\\

\noindent {\bf Acknowledgements}

I thank Dr. Hwayean Lee for her reading the original manuscript
and affirming the improvement of their work. This work is
supported by the National Natural Science Foundation of China
under Grant No.10304022, the science-technology fund of Anhui
province for outstanding youth under Grant No.06042087, the
general fund of the educational committee of Anhui province under
Grant No.2006KJ260B, and the key fund of the ministry of education
of China under Grant No.206063. \\

\noindent {\bf Reference}

\noindent[1] C. H. Bennett and G. Brassard, in {\it Proceeding of
the IEEE International Conference on Computers, Systems, and
Signal Processing Bangalore 1984} (IEEE, New York, 1984) pp175

\noindent[2] N. Gisin, G. Ribordy, W. Tittel, and H. Zbinden, Rev.
Mod. Phys. {\bf 74}, 145 (2002).

\noindent[3] A. Beige, B. G. Englert, C. Kurtsiefer, and
H.Weinfurter, Acta Phys. Pol. A {\bf101}, 357 (2002).

\noindent[4] K. Bostrom K and F. Felbinger. Phys. Rev. Lett.
{\bf89}, 187902 (2002).

\noindent[5] F. G. Deng, G. L. Long, and  X. S. Liu, Phys. Rev. A
{\bf68}, 042317 (2003).

\noindent[6] F. G. Deng and G. L. Long,  Phys. Rev. A {\bf69},
 052319 (2004).

\noindent[7] B. A. Nguyen,  Phys. Lett. A {\bf 328}, 6 (2004).

\noindent[8] Z. J. Zhang, Z. X. Man, and Y. Li, Int. J. Quantum
Information {\bf 2}, 521 (2004).

\noindent[9] Z. J. Zhang, Z. X. Man, and Y. Li, Chin. Phys. Lett.
{\bf 22}, 18 (2005).

\noindent[10] M. Lucamaini and S. Mancini, Phys. Rev. Lett.
{\bf94}, 140501 (2005).

\noindent[11] C. Wang, F. G. Deng, Y. S. Li, X. S. Liu, and G. L.
Long, Phys. Rev. A {\bf 71}, 044305 (2005).

\noindent[12] Hwayean Lee, Jongin Lim, and HyungJin Yang, Phys.
Rev. A {\bf 73}, 042305 (2006).

\noindent[13] Z. X. Man, Z. J. Zhang, and Y. Li, Chin. Phys. Lett.
{\bf 22}, 22 (2005).

\noindent[14] Z. J. Zhang, Y. Li, and Z. X. Man, Phys. Lett. A
{\bf 341}, 385 (2005).

\noindent[15] Z. J. Zhang, quant-ph/0604035.

\noindent[16] D. M. Greenberg, M. A. Horne, A. Shimony, and Z.
Zeilinger, Am. J. Phys. {\bf 58}, 1131 (1990).

\noindent[17] Z. J. Zhang, Y. Li, and Z. X. Man, Phys. Rev. A
{\bf71},  044301 (2005).

\noindent[18] Z. J. Zhang and Z. X. Man, Phys. Rev. A {\bf72},
022303 (2005).

\enddocument